\documentclass[preprint,12pt]{elsarticle}
\biboptions{sort&compress}
\usepackage{array}
\usepackage{longtable,booktabs}
\usepackage{amsmath,bbold,amssymb,epsfig,bm,mathrsfs}
\usepackage{feynmp,color,slashed,nicefrac,exscale,multirow,graphicx}
\usepackage{epstopdf}
\usepackage{longtable}
\usepackage{multirow}

\usepackage[dvipdfm,bookmarks=true,colorlinks,%
            citecolor=blue,linkcolor=blue,hypertex, %
            breaklinks=true]{hyperref}

\begin{document}

\begin{frontmatter}
%\begin{CJK*}{GBK}{song}
%\begin{CJK}{UTF8}{}
\hyphenpenalty=6000
\tolerance=1000

\title{Possible antimagnetic rotation bands in $^{100}$Pd:
       a particle-number conserving investigation}

\author{Jian-Qin Ma}
\author{Zhen-Hua Zhang\fnref{contact}}
\ead{zhzhang@ncepu.edu.cn}
\address{Mathematics and Physics Department,
         North China Electric Power University, Beijing 102206, China}

\begin{abstract}
The particle-number conserving method based on the cranked shell model
is adopted to investigate the possible antimagnetic rotation bands in $^{100}$Pd.
The experimental kinematic and dynamic moments of inertia,
together with the $B(E2)$ values are reproduced quite well.
The occupation probability of each neutron and proton orbital in the observed
antimagnetic rotation band is analyzed and its configuration is confirmed.
The contribution of each major shell to the total angular momentum
alignment with rotational frequency in the
lowest-lying positive and negative parity bands is analyzed.
The level crossing mechanism of these bands is understood clearly.
The possible antimagnetic rotation in the negative parity $\alpha=0$ branch
is predicted, which sensitively depends on the alignment of the
neutron ($1g_{7/2}$, $2d_{5/2}$) pseudo-spin partners.
The two-shears-like mechanism for this antimagnetic rotation is investigated
by examining the closing of the proton hole angular momentum vector
towards the neutron angular momentum vector.
\end{abstract}


\end{frontmatter}



\section{\label{Sec:Introduction}Introduction}

In the near spherical or weakly deformed nuclei, the total angular momentum
is almost fully generated by the valence nucleons instead of the collective rotation.
Antimagnetic rotation (AMR)~\cite{Frauendorf1996_Proceedings, Frauendorf2001_RMP73-463},
which is predicted by Frauendorf in analogy to the antiferromagnetism in solids,
is an interesting exotic phenomenon in high-spin physics.
The AMR bands have regular sequences of energy levels differing
in spin by $2\hbar$ and are connected by $E2$ transitions,
which are very similar to the traditional collective rotation.
However, the $E2$ transitions in AMR bands are very weak,
reflecting the weakly deformed or nearly spherical core.
In AMR bands, two paired nucleons in a two-shears configuration
are aligned back to back in opposite direction and nearly perpendicular
to the total angular momentum vector of the valence neutrons.
Angular momentum is generated by the two-shears-like mechanism,
i.e., by simultaneously closing of the angular momentum vector of two
valence protons toward the total neutron angular momentum vector.
One typical feather of AMR band is the decreasing of the $B(E2)$ values with
increasing spin, which can be confirmed by the lifetime measurement.

The experimental evidence on AMR is very scarce.
According to the lifetime measurement, AMR bands with different configurations
have been observed mainly in the weakly deformed Cd ($Z=48$) and Pd ($Z=46$)
isotopes in the $A \approx 100$ mass region,
including $^{105-108,110}$Cd~\cite{Choudhury2010_PRC82-061308R, Simons2003_PRL91-162501,
Choudhury2013_PRC87-034304, Simons2005_PRC72-024318, Datta2005_PRC71-041305R, Roy2011_PLB694-322}
and $^{101, 104}$Pd~\cite{Sugawara2012_PRC86-034326, Sugawara2015_PRC92-024309,
Singh2017_JPG44-075105, Rather2014_PRC89-061303R}.
In addition, several possible AMR bands have been observed in
$^{109}$Cd~\cite{Chiara2000_PRC61-034318},
$^{112}$In~\cite{Li2012_PRC86-057305},
$^{144}$Dy~\cite{Sugawara2009_PRC79-064321}, and
$^{142,143}$Eu~\cite{Ali2017_PRC96-021304R, Rajbanshi2015_PLB748-387},
which need further confirmation by lifetime measurements.

From the theoretical aspect, AMR has been investigated mainly by the
semi-classical particle rotor model~\cite{Clark2000_ARNPS50-1},
and the tilted axis cranking (TAC) model~\cite{Frauendorf2000_NPA677-115,
Peng2008_PRC78-024313, Zhao2011_PLB699-181, Meng2013_FP8-55, Zhao2017_PLB773-1}.
Based on the TAC model, many investigations have been performed within the framework of
pairing plus quadrupole model~\cite{Chiara2000_PRC61-034318, Frauendorf2001_RMP73-463},
microscopic-macroscopic model~\cite{Zhu2001_PRC64-041302R, Simons2003_PRL91-162501, Simons2005_PRC72-024318},
and the covariant density functional theory~\cite{Zhao2011_PRL107-122501, Zhao2012_PRC85-054310,
Liu2012_SSPMA55-2420, Zhang2014_PRC89-047302, Peng2015_PRC91-044329, Jia2018_PRC97-024335}
with the point coupling effective interaction PC-PK1~\cite{Zhao2010_PRC82-054319}.

Recently, the existence of AMR band has been confirmed in $^{100}$Pd~\cite{Sihotra2020_PRC102-034321}
by lifetime measurement for the yrast negative-parity band.
The high-spin states of this nucleus have already been investigated
successfully in Refs.~\cite{Perez2001_NPA686-41, Zhu2001_PRC64-041302R}.
For even-even nuclei, most of the AMR bands are the lowest-lying
positive parity bands (yrast bands) except $^{100}$Pd,
which is observed with negative parity.
This indicates that the observed AMR band in $^{100}$Pd is based on a two-quasiparticle configuration.
Ref.~\cite{Perez2001_NPA686-41} suggested that it is built up
from a two-quasineutron configuration $\nu^2 1/2^-[550](\alpha=-1/2)1/2^+[420](\alpha=-1/2)$.
If the rotational properties of the AMR band and its signature partner
are similar, two AMR bands can exist in $^{100}$Pd.
Note that the first multiple AMR bands have been observed
in $^{107}$Cd~\cite{Choudhury2013_PRC87-034304}, which are signature partners.
Therefore, it is interesting to investigate the possible AMR bands in $^{100}$Pd.

In the present work, the particle-number conserving (PNC)
method based on the cranked shell model (CSM)~\cite{Zeng1994_PRC50-1388, Zhang2020_PRC101-054303}
will be adopted to investigate the possible AMR bands in $^{100}$Pd.
Note that with exact particle-number conservation for treating the pairing correlations,
successful descriptions have already been archived for
the AMR bands in $^{105, 106}$Cd~\cite{Zhang2013_PRC87-054314} and
$^{101, 104}$Pd~\cite{Zhang2016_PRC94-034305, Zhang2019_ChinPhysC43-054107} by PNC-CSM.

This paper is organized as follows.
A brief introduction of PNC-CSM is presented in Sec.~\ref{Sec:PNC-CSM}.
The results and discussion about the possible AMR bands in $^{100}$Pd are given in Sec.~\ref{Sec:Results}.
Finally, the summary of this work is given in Sec.~\ref{Sec:Summary}.

\section{\label{Sec:PNC-CSM}Theoretical framework}

The cranked shell model Hamiltonian in the rotating frame reads
\begin{eqnarray}
 H_\mathrm{CSM} = H_0 + H_\mathrm{P} = H_{\rm Nil}-\omega J_x + H_\mathrm{P} \ ,
 \label{eq:H_CSM}
\end{eqnarray}
where $H_{\rm Nil}$ is the Nilsson Hamiltonian, $-\omega J_x$ is the
Coriolis interaction with the cranking frequency $\omega$ about the
$x$ axis (perpendicular to the nuclear symmetry $z$ axis).
$H_{\rm P}$ is the pairing interaction.
In the present investigation, monopole pairing
\begin{eqnarray}
 H_{\rm P} = -G \sum_{\xi\eta} a^\dag_{\xi} a^\dag_{\bar{\xi}}
             a_{\bar{\eta}} a_{\eta}  \ ,
\end{eqnarray}
is adopted, where $\bar{\xi}$ ($\bar{\eta}$) denotes the time-reversal state of the
Nilsson state $\xi$ ($\eta$), and $G$ is the effective monopole pairing strength.

Different from the single-particle level truncation in the conventional shell-model calculation,
a cranked many-particle configuration (CMPC) truncation is adopted in the PNC method,
which is important to make the shell-model like calculation
both feasible and sufficiently accurate~\cite{Wu1989_PRC39-666,Molique1997_PRC56-1795}.
Generally speaking, the dimension of CMPC space with 1000
is enough for both proton and neutron when investigating the heavy nuclei.

By diagonalizing the cranking Hamiltonian $H_\mathrm{CSM}$ in a sufficiently
large CMPC space, the wave-functions for the yrast and the
low-lying excited states can be obtained, which can be written as
\begin{equation}
 |\Psi\rangle = \sum_{i} C_i \left| i \right\rangle \ ,
 \qquad (C_i \; \textrm{real}),
\end{equation}
where $| i \rangle$ is a CMPC and $C_i$ is the corresponding coefficient.
The expectation value of a one-body operator
$\mathcal {O} = \sum_{k=1}^N \mathscr{O}(k)$ can be written as
\begin{equation}
 \left\langle \Psi | \mathcal {O} | \Psi \right\rangle
 =\sum_i C_i^2 \left\langle i | \mathcal {O} | i \right\rangle
 +2\sum_{i<j} C_i C_j \left\langle i | \mathcal {O} | j \right\rangle \ .
\end{equation}
The matrix element $\langle i | \mathcal {O} | j \rangle$ for $i\neq j$
is nonzero only when the two CMPCs $|i\rangle$ and $|j\rangle$ differ by
one particle occupation~\cite{Zeng1994_PRC50-1388}.
After a certain permutation of creation operators,
$|i\rangle$ and $|j\rangle$ can be rewritten as
\begin{equation}
 | i \rangle = (-1)^{M_{i\mu}} | \mu \cdots \rangle \ , \qquad
 | j \rangle = (-1)^{M_{j\nu}} | \nu \cdots \rangle \ ,
\end{equation}
where $\mu$ and $\nu$ denote two different single-particle states,
and $(-1)^{M_{i\mu}}=\pm1$, $(-1)^{M_{j\nu}}=\pm1$ according to
whether the permutation is even or odd.
Therefore, the expectation value of the one-body operator $\mathcal {O}$
can be separated into the diagonal $\sum_{\mu} \mathscr{O}(\mu)$
and the off-diagonal $2\sum_{\mu<\nu} \mathscr{O}(\mu\nu)$ parts
\begin{eqnarray}
 \mathcal {O}
  &=& \left\langle \Psi | \mathcal {O} | \Psi \right\rangle =
  \left( \sum_{\mu} \mathscr{O}(\mu) + 2\sum_{\mu<\nu} \mathscr{O}(\mu\nu) \right) \ , \label{eq:j1}\\
 \mathscr{O}(\mu)
 &=& \langle \mu | \mathscr{O} | \mu \rangle n_{\mu}  \ , \label{eq:j1d} \\
 \mathscr{O}(\mu\nu)
 &=&\langle \mu | \mathscr{O} | \nu \rangle
  \sum_{i<j} (-1)^{M_{i\mu}+M_{j\nu}} C_{i} C_{j} \ ,
  \label{eq:j1od}
\end{eqnarray}
where $n_{\mu}=\sum_{i}|C_{i}|^{2}P_{i\mu}$
is the occupation probability of the cranked single-particle orbital $|\mu\rangle$
and $P_{i\mu}=0$ (1) when $|\mu\rangle$ is empty (occupied) in the CMPC $|i\rangle$.
Note that in the traditional Bardeen-Cooper-Schrieffer or Hartree-Fock-Bogoliubov
approximations, the wave-functions are one Slatter determinant.
Therefore, there is only diagonal contribution in both of these two methods.

The kinematic moment of inertia (MOI) $J^{(1)}$ of the state $|\Psi\rangle$ is
\begin{eqnarray}
 J^{(1)} = \frac{1}{\omega} \left\langle \Psi | J_x | \Psi \right\rangle \ .
\end{eqnarray}

According to the semiclassical approximation,
the $B(E2)$ values can be obtained from
\begin{equation}
B(E2) = \frac{3}{8}
{\left\langle \Psi | Q_{20}^{\rm p} | \Psi \right\rangle}^2 \ ,
\end{equation}
where $Q_{20}^{\rm p}$ is the proton quadrupole moments and
\begin{equation}
Q_{20} = \sqrt{\frac{5}{16\pi}} (3z^2-r^2) = r^2 Y_{20} \ .
\end{equation}

\section{\label{Sec:Results} Results and discussion}

In the present work for $^{100}$Pd, Nilsson parameters ($\kappa$ and $\mu$)
are taken from the traditional values~\cite{Bengtsson1985_NPA436-14}.
The deformation parameters $\varepsilon_2 = 0.12$ and $\varepsilon_4=-0.02$
are taken from FRDM(2012)~\cite{Moller2016_ADNDT109-110-1}.
A smaller deformation $\varepsilon_2 = 0.083$ and $\varepsilon_4=-0.013$
taken from FRDM(1995)~\cite{Moeller1995_ADNDT59-185} is also adopted for comparison.
The valence single-particle space is constructed
from $N=0$ to $N=5$ major shells both for neutrons and protons.
Therefore, there is no effective charge involved when calculating the $B(E2)$ values.
The neutron and proton CMPC space are both chosen as 1000.
The pairing strengths are determined by the odd-even differences in nuclear
binding energy and adjusted slightly according to the MOIs at low rotational frequency region.
In the present work, $G_{\rm n} = G_{\rm p}$ = 0.5~MeV are adopted.

%%%%%%%%%%%%%%%%%%%%%%%%%%%%%%%%%%%%%%%%%%%%%%%%%%%%%%%%%%%%%%%%%%%%%%%%%%%%
\begin{figure}[h]
\includegraphics[width=0.8\columnwidth]{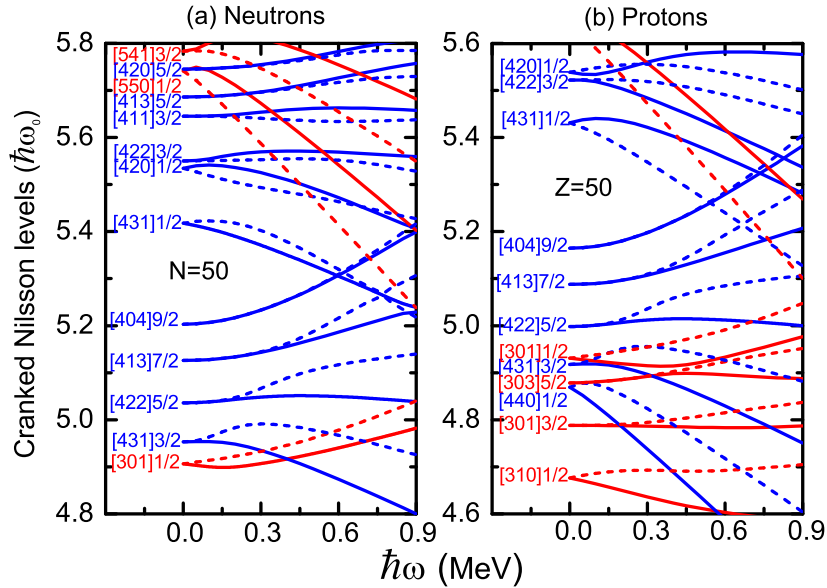}
\centering
\caption{\label{fig1:nil}
The cranked single-particle levels near the Fermi surface of $^{100}$Pd
for (a) neutrons and (b) protons.
The levels with positive and negative parity are displayed by blue and red lines, respectively.
The levels with signature $\alpha=+1/2$ and $\alpha=-1/2$
are displayed by solid and dashed lines, respectively.
}
\end{figure}
%%%%%%%%%%%%%%%%%%%%%%%%%%%%%%%%%%%%%%%%%%%%%%%%%%%%%%%%%%%%%%%%%%%%%%%%%%%%

The cranked single-particle levels near the Fermi surface of $^{100}$Pd
for neutrons and protons are given in Fig.~\ref{fig1:nil}.
The data show that the AMR band in $^{100}$Pd is the lowest-lying negative parity
band with signature $\alpha=1$~\cite{Sihotra2020_PRC102-034321}.
It can be seen in Fig.~\ref{fig1:nil}(a) that the lowest 2-quasineutron configuration
with $\alpha=1$ is $\nu^2 1/2^-[550](\alpha=-1/2)1/2^+[420](\alpha=-1/2)$,
which is consistent with the suggested configuration for this band in Ref.~\cite{Perez2001_NPA686-41}.
It can be seen from Fig.~\ref{fig1:nil}(b) that for the proton vacuum of $^{100}$Pd,
there are four $g_{9/2}$ holes ($\pi 7/2^+[413]$ and $\pi 9/2^+[404]$).
Due to the pairing correlations, they should be partly occupied at the bandhead.
Note that in the PNC calculation, specific orbitals are not blocked by hand.
Shell-model-like calculations are performed and the lowest-lying negative
parity band with $\alpha=1$ is assigned for the AMR band.
Its configuration can be obtained by the occupation probabilities of
the single-particle levels close to
the Fermi surface (see the discussions about Fig.~\ref{fig4:occup} below).

%%%%%%%%%%%%%%%%%%%%%%%%%%%%%%%%%%%%%%%%%%%%%%%%%%%%%%%%%%%%%%%%%%%%%%%%%%%%
\begin{figure}[h]
\includegraphics[width=0.8\columnwidth]{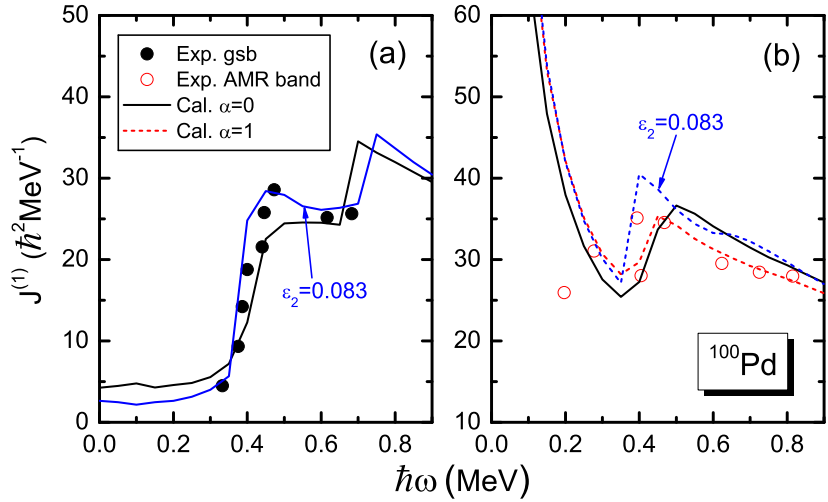}
\centering
\caption{\label{fig2:moi}
The experimental and calculated MOIs for (a) the ground state band
and (b) the AMR band in $^{100}$Pd.
The calculated results with the deformation ($\varepsilon_2=0.083, \varepsilon_4=-0.013$)
taken from FRDM(1995)~\cite{Moeller1995_ADNDT59-185}
are also shown as blue lines.
}
\end{figure}
%%%%%%%%%%%%%%%%%%%%%%%%%%%%%%%%%%%%%%%%%%%%%%%%%%%%%%%%%%%%%%%%%%%%%%%%%%%%

Figure~\ref{fig2:moi} shows the experimental and calculated MOIs for
the ground state band (gsb) and the AMR band in $^{100}$Pd.
The calculated results with a smaller deformation ($\varepsilon_2=0.083, \varepsilon_4=-0.013$)
taken from FRDM(1995)~\cite{Moeller1995_ADNDT59-185} are also shown.
It can be seen that the experimental MOIs can be reproduced quite well for both two bands
when adopting the deformation from FRDM(2012).
The upbendings in the gsb (two successive upbendings around $\hbar\omega\approx$~0.4~MeV)
and the AMR band (around $\hbar\omega\approx$~0.4~MeV) are also reproduced well.
In addition, an upbending in the gsb at $\hbar\omega\approx0.65$~MeV is predicted.
The level crossings in the gsb and the AMR band will be discussed later.
The signature partner of the AMR band
[$\nu^2 1/2^-[550](\alpha=-1/2)1/2^+[420](\alpha=+1/2)$] is also calculated.
It can be seen from Fig.~\ref{fig2:moi}(b) that its MOIs are quite different from the AMR band.
Compared with the AMR band, the MOIs are smaller at lower frequency region;
whereas at higher frequency region, the MOIs are larger,
which means that more alignments gain after the level crossing in this signature branch.
It also can be seen that if the smaller deformation [FRDM(1995)] is adopted,
the description of the gsb is improved,
while the description of the AMR band is getting worse.
This indicates that for the gsb, the deformation may decrease with spin
as suggested in Ref.~\cite{Perez2001_NPA686-41},
whereas the deformation of the AMR band may decrease not that much.

%%%%%%%%%%%%%%%%%%%%%%%%%%%%%%%%%%%%%%%%%%%%%%%%%%%%%%%%%%%%%%%%%%%%%%%%%%%%
\begin{figure}[!]
\includegraphics[width=0.6\columnwidth]{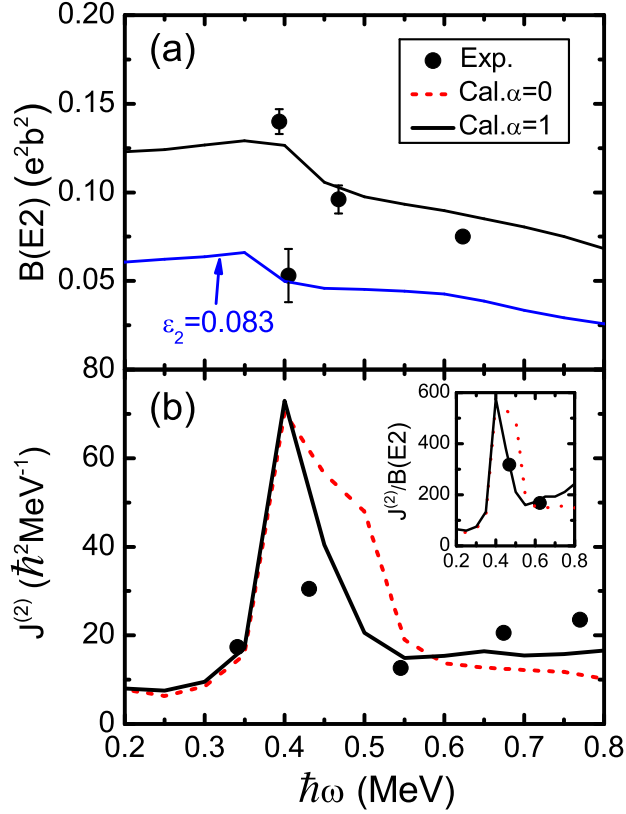}
\centering
\caption{\label{fig3:be2}
The experimental and calculated (a) $B(E2)$ values and (b) dynamic MOIs $J^{(2)}$
for the AMR band and its signature partner band (red dashed lines) in $^{100}$Pd.
The calculated $B(E2)$ values with the deformation ($\varepsilon_2=0.083, \varepsilon_4=-0.013$)
taken from FRDM(1995)~\cite{Moeller1995_ADNDT59-185} are shown as blue line.
The $J^{(2)}/B(E2)$ ratios are shown as an inset.
}
\end{figure}
%%%%%%%%%%%%%%%%%%%%%%%%%%%%%%%%%%%%%%%%%%%%%%%%%%%%%%%%%%%%%%%%%%%%%%%%%%%%

The typical feather of AMR band is the decreasing
of the $B(E2)$ values with increasing spin.
Fig.~\ref{fig3:be2}(a) shows the experimental and calculated $B(E2)$ values
for the AMR band in $^{100}$Pd.
Note that in the present PNC-CSM calculations, proton-neutron interaction is neglected.
Therefore, the $B(E2)$ values, which come from the contribution of proton subsystem,
are the same for the AMR band and its signature partner band.
It can be seen that the $B(E2)$ values are reproduced by the PNC-CSM and the decreasing
of the $B(E2)$ values with increasing rotational frequency can also be obtained.
The calculated results with the deformation from FRDM(1995) underestimate the
$B(E2)$ values a lot.
Since the $B(E2)$ values are very sensitive to the deformation,
this means that the deformation from FRDM(2012) is reasonable.
The adopted deformation is also consistent with
the Total Routhian Surface calculations in Ref.~\cite{Perez2001_NPA686-41}.
In principle, the deformation usually decreases with spin in the AMR band.
However, the PNC-CSM calculations show that even with the fixed
deformation, the $B(E2)$ values can still be reproduced quite well.
This indicates that the deformation may decrease not so much with spin in the AMR band of $^{100}$Pd.
Fig.~\ref{fig3:be2}(b) shows the comparison of experimental and calculated
dynamic MOIs $J^{(2)}$ for the AMR band and its signature partner band in $^{100}$Pd.
It can be seen that both the $J^{(2)}$ and $J^{(2)}/B(E2)$ ratios can be reproduced very well.
At $\hbar\omega >$ 0.55~MeV, the $J^{(2)}/B(E2)$ ratios increase with rotational frequency,
which is also a very most important feature of AMR.
The rise of $J^{(2)}/B(E2)$ ratios reflects that along the band, $J^{(2)}$
keeps nearly constant whereas the $B(E2)$ values decrease.
This is different from the behavior of smoothly terminating band~\cite{Afanasjev1999_PR322-1},
which has almost constant $J^{(2)}/B(E2)$ values,
indicating that the $J^{(2)}$ and $B(E2)$ values have
similar falling behavior with increasing rotational frequency.
The $J^{(2)}/B(E2)$ ratios of the $\alpha=0$ branch are also quite large,
which supports the possibility of the AMR in this signature branch.
The peak in the $J^{(2)}/B(E2)$ ratios in this signature branch around $\hbar\omega >$ 0.5~MeV
is due to the extra alignment, which will be discussed later.

%%%%%%%%%%%%%%%%%%%%%%%%%%%%%%%%%%%%%%%%%%%%%%%%%%%%%%%%%%%%%%%%%%%%%%%%%%%%
\begin{figure}[h]
\includegraphics[width=0.6\columnwidth]{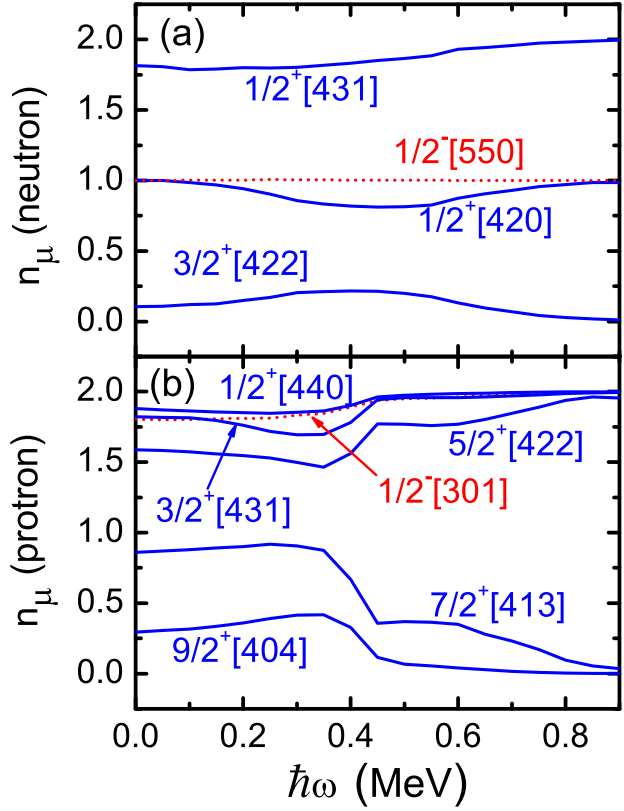}
\centering
\caption{\label{fig4:occup}
(a) Neutron and (b) proton occupation probability $n_\mu$ of each orbital
(including both signature $\alpha=\pm1/2$) close to the Fermi surface
of the AMR band in $^{100}$Pd.
}
\end{figure}
%%%%%%%%%%%%%%%%%%%%%%%%%%%%%%%%%%%%%%%%%%%%%%%%%%%%%%%%%%%%%%%%%%%%%%%%%%%%

Figure~\ref{fig4:occup} shows the occupation probability $n_\mu$ of each orbital
close to the Fermi surface of the AMR band in $^{100}$Pd.
It can be seen from Fig.~\ref{fig4:occup}(a) that the occupation probability of
$\nu1/2^-[550]$ is about 1 and the occupation probability of $\nu1/2^+[420]$ is very close to 1.
This means that the configuration of the AMR band is $\nu^2 1/2^-[550]1/2^+[420]$.
Further investigation shows that the signature of these two occupied levels is $\alpha=-1/2$.
Therefore, the configuration of the AMR band obtained by the present PNC-CSM
is consistent with that suggested in Ref.~\cite{Perez2001_NPA686-41}.
It should be noted that since $\nu 1/2^+[420](1g_{7/2})$
and $\nu 3/2^+[422](2d_{5/2})$ are pseudo-spin partners,
their occupation probabilities are mixed at the medium
rotational frequency region in the present calculation.
It can be seen from Fig.~\ref{fig4:occup}(b) that around $\hbar\omega >$ 0.4~MeV,
the occupation probabilities of two proton $g_{9/2}$ orbitals
$\pi 7/2^+[413]$ and $\pi 9/2^+[404]$ drop down quickly,
while the occupation probabilities for other proton $g_{9/2}$ orbitals,
e.g., $\pi 5/2^+[422]$ and $\pi 3/2^+[431]$, increase quickly.
This means that the first level crossing in the AMR band may be caused by the $g_{9/2}$ protons.
After the level crossing, the $\pi 9/2^+[404]$ is nearly empty
but the $\pi 7/2^+[413]$ is only partly empty.
In addition, the $\pi 5/2^+[422]$ is also not fully occupied.
Experimental data show that the AMR appears after the upbending
in the lowest-lying negative parity band with $\alpha=1$.
The present calculations show that the proton configuration of this AMR band is
four partly empty proton holes in $g_{9/2}$ orbital
due to the existence of pairing correlations.

%%%%%%%%%%%%%%%%%%%%%%%%%%%%%%%%%%%%%%%%%%%%%%%%%%%%%%%%%%%%%%%%%%%%%%%%%%%%
\begin{figure}[h]
\includegraphics[width=0.8\columnwidth]{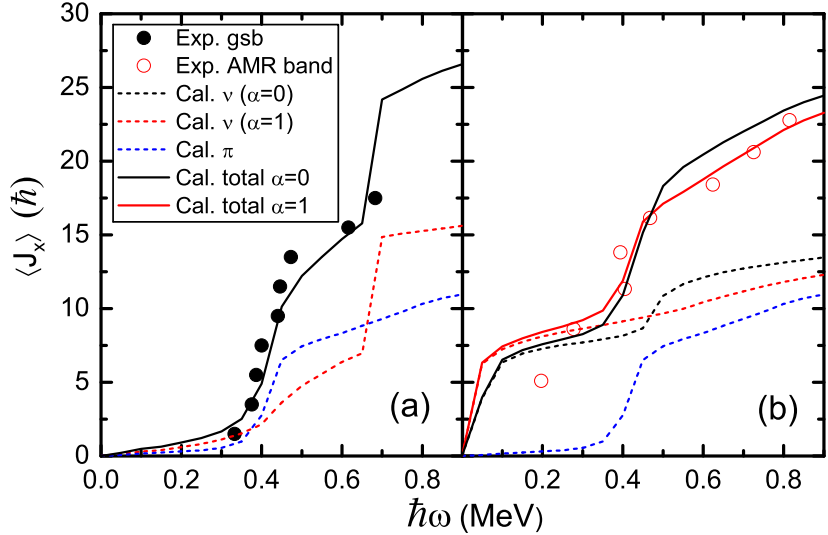}
\centering
\caption{\label{fig5:jx}
The experimental and calculated angular momentum alignments $\langle J_x\rangle$ for
(a) the gsb and (b) the AMR band in $^{100}$Pd.
The contributions from neutrons and protons are shown by dashed lines.
}
\end{figure}
%%%%%%%%%%%%%%%%%%%%%%%%%%%%%%%%%%%%%%%%%%%%%%%%%%%%%%%%%%%%%%%%%%%%%%%%%%%%

To investigate the level crossings in the gsb and the AMR band in $^{100}$Pd,
the experimental and calculated angular momentum alignments $\langle J_x\rangle$
are shown in Fig.~\ref{fig5:jx}.
It can be seen that the first level crossing in the gsb comes from the protons.
Just above this abrupt alignment, the calculations show a gradual alignment of neutrons.
This corresponds to the observed two successive upbendings in the experimental data.
The predicted sharp upbending at $\hbar\omega\approx0.65$~MeV comes from the contribution of neutron.
Fig.~\ref{fig5:jx}(b) shows that for the AMR band,
the sharp upbending comes from the contribution of proton just as the gsb.
For the signature $\alpha=0$ branch, a neutron alignment appears after the proton alignment.
This indicates that the AMR in this branch may also appear after this neutron alignment.
It should be noted that the level crossing frequency is sensitive to the
single-particle level scheme and also depends on the treatment of pairing correlations.
Since both the gsb and the AMR band are reproduced quite well in the present
calculation, the predicted neutron alignment in the $\alpha=0$ branch with
negative parity is reasonable.

%%%%%%%%%%%%%%%%%%%%%%%%%%%%%%%%%%%%%%%%%%%%%%%%%%%%%%%%%%%%%%%%%%%%%%%%%%%%
\begin{figure}[!]
\includegraphics[width=1.0\textwidth]{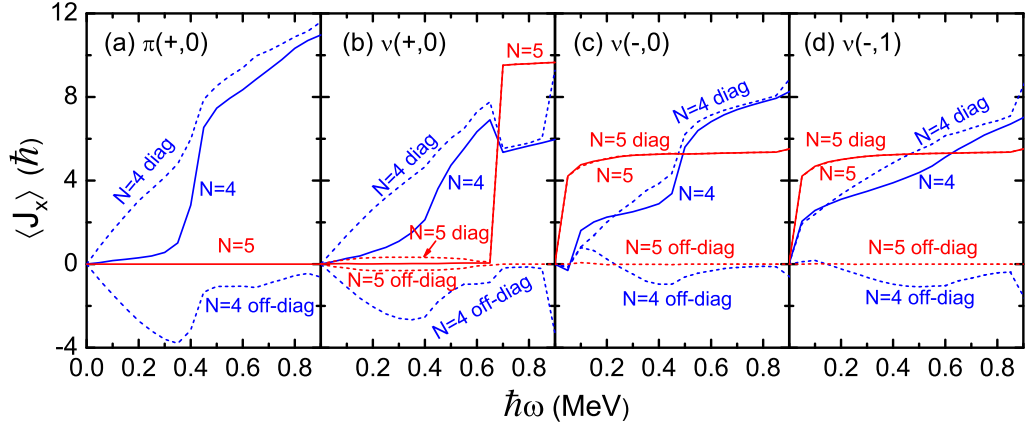}
\centering
\caption{\label{fig6:jxshell}
Contributions from $N=4$ and 5 major shells in the lowest-lying states of
(a) protons with positive parity and signature $\alpha=0$
(b) neutrons with positive parity and signature $\alpha=0$,
(c) neutrons with negative parity and signature $\alpha=0$, and
(d) neutrons with negative parity and signature $\alpha=1$,
to the total angular momentum alignment $\langle J_x\rangle$.
Contributions of diagonal and off-diagonal part are shown by dashed lines.
}
\end{figure}
%%%%%%%%%%%%%%%%%%%%%%%%%%%%%%%%%%%%%%%%%%%%%%%%%%%%%%%%%%%%%%%%%%%%%%%%%%%%

To see the alignment process in the gsb and the AMR band more clearly,
contributions from neutron and proton $N=4$ and 5 major shells to the total angular
momentum alignment $\langle J_x\rangle$ are shown in Fig.~\ref{fig6:jxshell}.
It can be seen in Fig.~\ref{fig6:jxshell}(a) that, the sharp upbending
at $\hbar\omega\approx0.4$~MeV in the gsb mainly comes from the proton
$N=4$ major shell, in which both of the diagonal and off-diagonal parts have contributions.
Fig.~\ref{fig6:jxshell}(b) shows that the successive
second gradual upbending around $\hbar\omega\approx$~0.45~MeV
in the gsb mainly comes from the off-diagonal part of neutron $N=4$ major shell.
If we look into detail, it comes from the
interference between the ($1g_{7/2}$, $2d_{5/2}$) pseudo-spin partners.
In addition, the upbending predicted at $\hbar\omega\approx0.65$~MeV
is caused by the diagonal part of neutron $N=5$ major shell, i.e.,
the neutron $h_{11/2}$ orbitals.
This indicates that the AMR can't exist in the gsb of $^{100}$Pd.
Fig.~\ref{fig6:jxshell}(c) shows that neutron $N=4$ shell contributes to
the upbending in the signature $\alpha=0$ negative parity band
just after the proton alignment.
This upbending is also caused by the neutron ($1g_{7/2}$, $2d_{5/2}$) pseudo-spin partners.
Different from the positive parity branch, both diagonal and off-diagonal
parts of these orbitals contribute to this upbending.

%%%%%%%%%%%%%%%%%%%%%%%%%%%%%%%%%%%%%%%%%%%%%%%%%%%%%%%%%%%%%%%%%%%%%%%%%%%%
\begin{figure}[h]
\includegraphics[width=0.6\columnwidth]{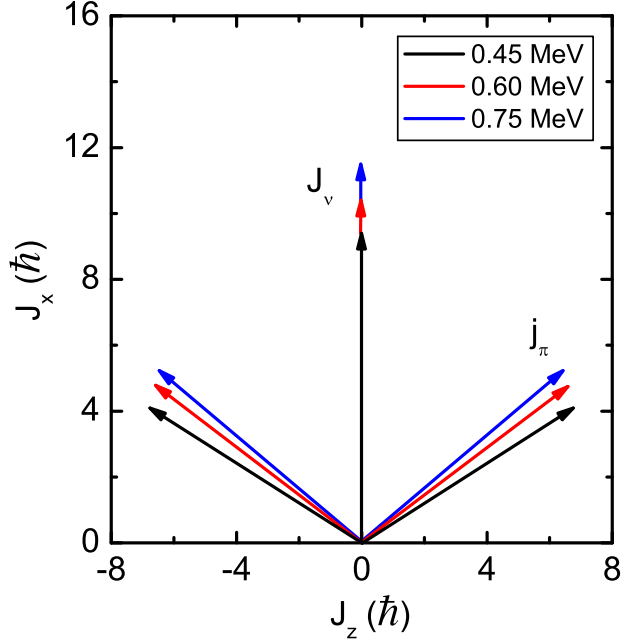}
\centering
\caption{\label{fig7:shear}
Angular momentum vectors of neutrons ($J_\nu$) and
four $g_{9/2}$ proton holes ($j_\pi$) for the AMR band in $^{100}$Pd.
Each $j_\pi$ includes the contribution from two $g_{9/2}$ proton holes.
}
\end{figure}
%%%%%%%%%%%%%%%%%%%%%%%%%%%%%%%%%%%%%%%%%%%%%%%%%%%%%%%%%%%%%%%%%%%%%%%%%%%%

Figure~\ref{fig7:shear} shows the angular momentum vectors of neutrons ($J_\nu$) and
four $g_{9/2}$ proton holes ($j_\pi$) for the AMR band in $^{100}$Pd
at rotational frequencies from 0.45 to 0.75~MeV.
Each $j_\pi$ includes the contribution from two $g_{9/2}$ proton holes.
It should be noted that these four proton holes are partly
occupied in the $g_{9/2}$ orbital due to the
pairing correlations [see Fig.~\ref{fig4:occup}(b)].
In the present PNC-CSM formalism principal axis cranking is assumed,
and $J_z$ is calculated approximately by
\begin{equation}
J_z = \sqrt{\langle \Psi | J_z^2 | \Psi \rangle}
\end{equation}
according to Ref.~\cite{Frauendorf1996_ZPA356-263}.
One can see from Fig.~\ref{fig7:shear} that with rotational frequency increasing,
the two proton $j_\pi$ vectors gradually close toward the neutron $J_\nu$ vector,
while the direction of the total angular momentum stays unchanged.
Therefore, higher angular momentum can be generated by the two-shears-like mechanism.
Note that the angular momentum contributed from $g_{9/2}$ proton holes
is only a little larger than that from the neutron, which comes from the collective rotation.
In addition, by analyzing the MOIs, we obtained that
the deformation decreases not so much with spin in this AMR band.
Therefore, the observed AMR band in $^{100}$Pd is not pure but highly mixed with the common rotation.
This is quite different from the neighboring nuclei $^{101}$Pd,
in which the rearrangement of occupations in the proton $g_{9/2}$ holes plays
an important role in the two-shears-like
mechanism~\cite{Zhang2016_PRC94-034305, Liu2019_PRC99-024317}.
The two-shears-like mechanism for the predicted AMR band (signature $\alpha=0$ branch)
is the same as that in Fig.~\ref{fig7:shear} for the proton blade,
and only a little different in the neutron momentum vector.
So we only show the two-shears-like mechanism for the observed AMR band as an example.

\section{\label{Sec:Summary}Summary}

In summary, the particle-number conserving method based on the cranked shell model
is adopted to investigate the possible antimagnetic rotation bands in $^{100}$Pd.
In the present calculations, the particle-number is conserved
and the Pauli blocking effects are treat exactly.
The experimental kinematic and dynamic moments of inertia,
together with the $B(E2)$ values are reproduced very well.
The investigation reveals that for the ground state band,
the deformation may decrease with spin,
whereas the deformation of the antimagnetic rotation band may decrease not so much.
The occupation probability of each neutron and proton orbital in the
antimagnetic rotation band is analyzed and its configuration is confirmed.
The contribution of each major shell to the total angular momentum alignment with
rotational frequency in the lowest-lying positive and negative parity bands is analyzed,
and their level crossing mechanism is understood clearly.
A sharp upbending is predicted at $\hbar\omega\approx0.65$~MeV
in the ground state band, which is caused by the neutron $h_{11/2}$ orbitals.
Investigations indicate that the antimagnetic rotation can't exist in the ground state band.
The possible antimagnetic rotation in the negative parity $\alpha=0$ branch
is predicted, which sensitively depends on the alignment of the
neutron ($1g_{7/2}$, $2d_{5/2}$) pseudo-spin partners.
At last, the two-shears-like mechanism for this antimagnetic rotation is investigated
by examining the closing of the proton hole angular momentum vector
towards the neutron angular momentum vector.
It is found that the observed AMR band in $^{100}$Pd is highly mixed with the common rotation.

\section*{Acknowledgement}

This work is supported by National Natural Science Foundation of
China (No. 11875027, 11775112, 11775026, 11775099, 11975096),
and the Fundamental Research Funds for the Central Universities (2021MS046).

%\bibliography{../../../../../Refecences/Papers}

%

\end{document}